\documentclass[11pt]{article}
\usepackage{geometry}
\usepackage{times}  % or \usepackage{lmodern} for improved fonts
\usepackage[utf8]{inputenc}

\usepackage{amsmath}
\usepackage{amssymb}
\usepackage{algorithm}
\usepackage{algorithmicx}
\usepackage{algpseudocode}
\usepackage{stfloats}
\usepackage{hyperref}
\usepackage{comment}

\usepackage{amsthm}
\usepackage{graphicx}
\usepackage{xcolor}
\usepackage{subcaption}
\usepackage{enumitem}
\usepackage{adjustbox}
\usepackage{paralist}
\usepackage{array}
\usepackage[switch]{lineno}
\newtheorem{definition}{Definition}
\newcommand{\model}{ConLES}
\newcommand{\modelspace}{ConLES }

\newcommand{\Input}{\Require}
\newcommand{\Output}{\Ensure}
\usepackage{adjustbox}
\title{Conformance Checking for Less: Efficient Conformance Checking for Long Event Sequences}
\date{}
\author{
Eli Bogdanov 
\and
Izack Cohen 
\and
Avigdor Gal 
}

\begin{document}
\maketitle
%\begin{CCSXML}
%<ccs2012>
% <concept>
%  <concept_id>00000000.0000000.0000000</concept_id>
%  <concept_desc>Do Not Use This Code, Generate the Correct Terms for Your Paper</concept_desc>
%  <concept_significance>500</concept_significance>
% </concept>
% <concept>
%  <concept_id>00000000.00000000.00000000</concept_id>
%  <concept_desc>Do Not Use This Code, Generate the Correct Terms for Your Paper</concept_desc>
%  <concept_significance>300</concept_significance>
% </concept>
% <concept>
%  <concept_id>00000000.00000000.00000000</concept_id>
%  <concept_desc>Do Not Use This Code, Generate the Correct Terms for Your Paper</concept_desc>
%  <concept_significance>100</concept_significance>
% </concept>
% <concept>
%  <concept_id>00000000.00000000.00000000</concept_id>
%  <concept_desc>Do Not Use This Code, Generate the Correct Terms for Your Paper</concept_desc>
%  <concept_significance>100</concept_significance>
% </concept>
%</ccs2012>
%\end{CCSXML}

%\ccsdesc[500]{General and reference~Metrics}
%\ccsdesc[500]{Information systems~Process mining}
%\ccsdesc[300]{Theory of computation~Formal languages and automata theory}
\begin{abstract}
Long event sequences (termed {\em traces}) and large data logs that originate from sensors and prediction models are becoming increasingly common in our data-rich world. In such scenarios, conformance checking—validating a data log against an expected system behavior (the {\em process model})—can become computationally infeasible due to the exponential complexity of finding an optimal alignment.
To alleviate scalability challenges for this task, we propose \model, a sliding-window conformance checking approach for long event sequences that preserves the interpretability of alignment-based methods. \modelspace partitions traces into manageable subtraces and iteratively aligns each against the expected behavior, leading to significant reduction of the search space while maintaining overall accuracy.
We use global information that captures structural properties of both the trace and the process model, enabling informed alignment decisions and discarding unpromising alignments, even if they appear locally optimal.
Performance evaluations across multiple datasets highlight that \modelspace outperforms the leading optimal and heuristic algorithms for long traces, consistently achieving the optimal or near-optimal solution. Unlike other conformance methods that struggle with long event sequences, \modelspace significantly reduces the search space, scales efficiently, and uniquely supports both predefined and discovered process models, making it a viable and leading option for conformance checking of long event sequences.
\end{abstract}

%\keywords{Conformance checking, Process mining, Long event sequences, Sliding window}

% \noindent {\bf Code, Datasets, Extended version --}\\ \url{https://github.com/runninghorizon/runninghorizon}

\section{Introduction}\label{sec:intro}
Widespread usage of sensors, the Internet of Things (IoT), and prediction models in industries such as construction and aerospace led to an abundance of data, generating long event sequences (termed {\em traces}) with thousands of events~\cite{beerepoot2023biggest, bogdanov2022conformance, bogdanov2023sktr}. 
The task of aligning such traces with normative system processes (termed {\em conformance checking}) has been identified as a leading contemporary challenge~\cite{beerepoot2023biggest}. Conformance checking ensures accurate process performance by identifying areas for improvement while ensuring customer satisfaction by adherence to predetermined criteria.

Conformance checking over long event sequences poses a challenge to contemporary methods, preventing them from successfully completing the task (see our empirical analysis in Section~\ref{sec:empirical_evaluation}). 
By breaking a long trace into manageable subtraces and applying a sliding window technique, we present a scalable, efficient solution. The process model retains its state between subtrace alignments, ensuring continuous and coherent analysis. The proposed technique can be adjusted using user-defined hyperparameters to balance computational efficiency and alignment accuracy.

The proposed approach leverages global information to make informed subtrace-level decisions. It calculates a lower bound on the marginal cost of a subtrace by considering the number of remaining activities that cannot be executed from each model state, adjusting this cost based on the frequency of unreachable transitions in the remaining trace.

Performance evaluations across multiple datasets demonstrate that it consistently finds optimal or near-optimal solutions. Our experiments show that this approach can handle traces with thousands of events - surpassing the capabilities of other leading methods—while maintaining reasonable running times.

The main contributions of this work are as follows:
\begin{compactitem}
    \item \textbf{Modeling and algorithm development:} We introduce \model, a novel approach that decomposes long traces while maintaining the process model intact. We demonstrate how segmentation enables efficient handling of event logs without compromising the overall process analysis integrity. The algorithm accounts for various nuances to efficiently maintain and evaluate multiple alignment paths within the process graph.
    \item \textbf{Demonstration of scalability:} Complexity analyses of the developed algorithm and evaluations demonstrate its superior scalability compared to leading alternatives.
    \item \textbf{Empirical validation:} Experiments on both known and novel benchmark datasets with long traces demonstrate \model~efficiency and effectiveness. The results indicate significantly reduced computational overhead with near-optimal alignment accuracy compared to optimal methods, and substantially higher conformance accuracy with comparable or improved runtime compared to state-of-the-art heuristic approaches.
\end{compactitem}
 The rest of the paper is organized as follows. We present the basics of alignment-based conformance checking (Section~\ref{sec:model}), followed by the introduction of \modelspace (Section~\ref{sec:method}). A formal complexity analysis (Section~\ref{sec:complexity}) is followed by an empirical evaluation (Section~\ref{sec:empirical_evaluation}), related work overview (Section~\ref{sec:literature_review}) and concluding remarks (Section~\ref{sec:conclusions_and_future_directions}).
 
\section{Preliminaries}\label{sec:model}

We tackle an alignment-based conformance checking problem~\cite{adriansyah2014aligning}, which aims to identify discrepancies between process executions (traces) and an expected behavior formalized as a process model. To this end, we introduce the foundational concepts of our approach: Petri nets, trace models, synchronous products, cost functions, and optimal alignments~\cite{carmona2018conformance}.

\begin{figure}[h!]
	\centering
	\includegraphics[width=0.95\columnwidth]{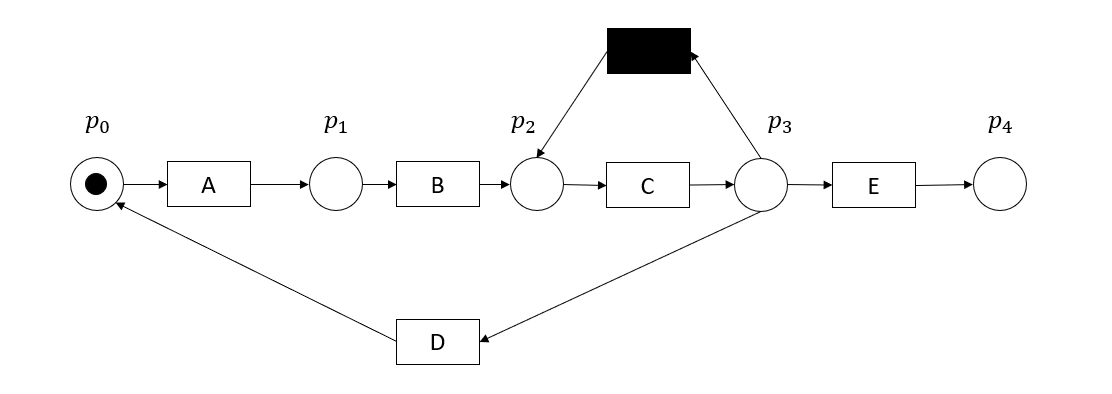}
	\caption{An example of a process model, showing the sequence of activities and their relationships.}
	%\Description{A process model diagram depicting a workflow with nodes representing activities and edges representing transitions between them.The model includes start and end points, decision points, and parallel execution paths.}
	\label{fig:processmodelrunningexample}
\end{figure}

Processes are modeled using \emph{labeled Petri nets}, which provide a formal representation of concurrent activities and complex synchronization mechanisms. A labeled Petri net consists of two types of nodes: \emph{places}, depicted as circles, and \emph{transitions}, represented as rectangles. Directed arcs connect places and transitions, alternating between the two types. Figure~\ref{fig:processmodelrunningexample} illustrates a labeled Petri net with six transitions, including a silent transition ($\tau$, marked as a black rectangle) that represents an internal orchestration step rather than a visible activity.
Tokens, depicted as black dots, reside in places and flow through the arcs and transitions. The distribution of tokens across places reflects the current state (or \emph{marking}) of the Petri net. A valid sequence of token flows from the initial marking to a final marking represents a possible execution of the process.

\begin{definition}[Labeled Petri Net] \label{def:labelled_petri_net}
Let \( \mathcal{A} \) be the universe of all possible activities, and let \( A \subseteq \mathcal{A} \) be a set of activities relevant to a specific process or trace. A labeled Petri net is a tuple \( N = (P, T, F, \lambda) \), where:
\begin{compactitem}
    \item \( P \) is a finite set of \emph{places},
    \item \( T \) is a finite set of \emph{transitions}, with \( P \cap T = \emptyset \),
    \item \( F \subseteq (P \times T) \cup (T \times P) \) represents a set of \emph{flow relations} as directed arcs between places and transitions,
    \item \( \lambda: T \to A \cup \{\tau\} \) is a \emph{labeling function} assigning to transitions \( t \in T \) either an activity label from \( A \) or the special symbol \( \tau \). %(indicating a silent transition not associated with any activity).
\end{compactitem}
\end{definition}

\begin{definition}[Trace Model] \label{def:trace_model}
Let \( A \subseteq \mathcal{A} \) be a set of activities over the universe of all possible activities \( \mathcal{A} \), and let \( \sigma \in A^* \) be a sequence from the set of all possible activity sequences. A trace model is a tuple \( TN = (P, T, F, \lambda, m_i, m_f) \), where:
\begin{compactitem}
    \item \( P = \{p_0, \ldots, p_{|\sigma|}\} \) is a finite set of \emph{places},
    \item \( T = \{t_1, \ldots, t_{|\sigma|}\} \) is a finite set of \emph{transitions},
    \item \( F = \{(p_i, t_{i+1}) \mid 0 \leq i < |\sigma|\} \cup \{(t_i, p_i) \mid 1 \leq i \leq |\sigma|\} \) is a set of \emph{flow relations},
    \item \( m_i = [p_0] \) is the \emph{initial marking}, assigning one token to the initial place \( p_0 \),
    \item \( m_f = [p_{|\sigma|}] \) is the \emph{final marking}, assigning one token to the final place \( p_{|\sigma|} \),
    \item \( \lambda: T \to A \) is a \emph{labeling function} that assigns to each transition \( t_i \in T \) the corresponding activity \( \sigma(i) \).
\end{compactitem}
\end{definition}

\begin{figure}[h!]
	\centering
	\includegraphics[width=0.95\columnwidth]{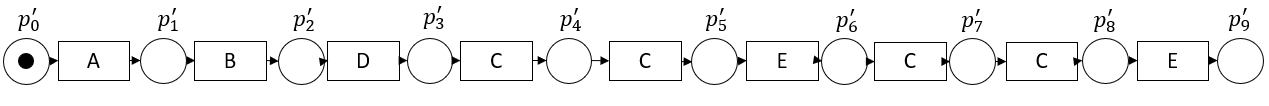}
	\caption{An example of a trace model, illustrating the execution path of a specific process instance.}
	%\Description{	A trace model diagram showing the sequence of events or activities executed in a specific process instance.The trace includes timestamps, activity labels, and transitions between activities.}
	\label{fig:tracemodelrunningexample}
\end{figure}

Definition~\ref{def:trace_model} describes a trace as a special type of a Petri net. Figure~\ref{fig:tracemodelrunningexample} illustrates this concept using the trace $\langle ABDCCECCE \rangle$, a sequence of 9 transitions. 

Next, we define a synchronous product, a key concept for modeling the joint execution of a trace and a process model. %Using Petri net semantics and search techniques, it identifies matches and mismatches between a trace and a model.

\begin{sloppypar}
\begin{definition}[Synchronous Product]\label{def:sync_prod}
Let $A \subseteq \mathcal{A}$ be a set of activities,  $SN=(P^{SN},T^{SN},F^{SN},\lambda^{SN}, m_i^{SN}, m_f^{SN})$ a process model, and $\sigma \in A^*$  a trace with its corresponding trace model  $TN=(P^{TN},T^{TN},F^{TN},\lambda^{TN},m_i^{TN}, m_f^{TN})$. The synchronous product is a tuple $SP=(P,T,F,\lambda,m_i, m_f)$ such that:
\begin{compactitem}
    \item[$\bullet$] $P=P^{SN} \cup P^{TN}$ is the set of places.
    \item[$\bullet$] $T=T^{MM} \cup T^{LM}\cup T^{SM} \subseteq (T^{SN} \cup \{\!\gg\!\}) \times (T^{TN} \cup \{\!\gg\!\})$ is the set of transitions where $\gg$ denotes a new element, i.e., $\gg \notin T^{SN} \cup T^{TN}$, with \\
    \hspace*{2mm} $T^{MM} = T^{SN} \times \{\gg\}$ is the set of model moves, \\
    \hspace*{2mm} $T^{LM} = \{\gg\} \times T^{TN}$ is the set of log moves, and \\
    \hspace*{2mm} $T^{SM} \!=\! \{(t_1,t_2)\! \in\! T^{SN}\! \times T^{TN} \!\;\! | \!\;\! \lambda^{SN}(t_1) \!= \!\lambda^{TN}(t_2)\}$ is the set of synchronous moves.
    \item[$\bullet$] $F\!=\!\{(p,(t_1,t_2))\! \in P \!\times\! T \;\! | \; \!(p,t_1)\! \in \!F^{SN}\! \lor\! (p,t_2)\! \in \!F^{TN}\} \!\cup\! \{((t_1,t_2),p)\! \in\! T \!\times\! P \!\;|\; \!(t_1,p)\! \in \!F^{SN} \!\lor\! (t_2,p)\! \in\! F^{TN}\}$ is the set of flows,
    \item[$\bullet$] $m_i = m_i^{SN} + m_i^{TN}$ is the initial marking, 
    \item[$\bullet$] $m_f = m_f^{SN} + m_f^{TN}$ is the final marking, and 
    \item[$\bullet$] $\forall (t_1,t_2) \in T$ holds that $\lambda((t_1,t_2)) = (l_1,l_2)$, where $l_1 = \lambda^{SN}(t_1)$ if $t_1 \in T^{SN}$, and $l_1= \gg$ otherwise; and $l_2=\lambda^{TN}(t_2)$, if $t_2 \in T^{TN}$, and $l_2 = \gg$ otherwise. 
\end{compactitem}
\end{definition}
\end{sloppypar}

Cost functions play a critical role in alignment-based conformance checking, assigning numerical values to deviations detected through the synchronous product, like skipped activities or unexpected events. They quantify trace-process conformance by aggregating the costs of aligning trace events with process activities. The cost function that we use assigns a cost of 0 to synchronous moves, $\epsilon \to 0^+$~\cite{carmona2018conformance} to $\tau$-labeled moves, and a cost of 1 to other model and log moves. 

\begin{definition}[Cost Function] \label{def:cost_func}
Let \( SP = (P, T, F, \lambda, m_i, m_f) \) be a synchronous product. A cost function \( c: T \to \mathbb{R}^+ \cup \{0\} \) associates a non-negative cost with each transition firing. Specifically:
\begin{compactitem}
    \item \( c(t) = 0 \) for synchronous moves \( t \in T^{SM} \),
    \item \( c(t) = \epsilon \) (where \( \epsilon \to 0^+ \)) for silent moves \( t \in T^{MM} \) labeled with \( \tau \),
    \item \( c(t) = 1 \) for all other moves \( t \in T^{MM} \cup T^{LM} \).
\end{compactitem}
\end{definition}

In conformance checking, we seek an optimal alignment using the synchronous product and its corresponding reachability graph~\cite{carmona2018conformance}. This alignment, associated with the lowest cost as defined by the cost function, indicates the closest match between the trace and the process model, assessing the conformance of the process execution.

\begin{definition}[Optimal Alignment] \label{def:opt_alignment}
Let $A \subseteq \mathcal{A}$ be a set of activities, $\sigma \in A^*$ is a trace with $TN$ as its corresponding trace model, $SN$ a process model, and $SP$ the synchronous product of $SN$ and $TN$. Let $ c : T \rightarrow \mathbb{R}^+ \cup \{0\}$ be a cost function. An optimal alignment $\gamma^{opt} \in L_{SP}$, where $L_{SP}$ is the set of possible execution sequences, is a full execution sequence of the synchronous product such that for all $\gamma \in L_{SP}$, it holds that $c(\gamma) \geq c(\gamma^{opt})$, where $c(\gamma) = \sum_{1\leq i \leq  |\gamma|} \, c(\gamma(i))$.
\end{definition}

An optimal alignment can be found using $A^*$-based methods over the reachability graph of a synchronous product, progressing from initial to final markings. At each marking, the algorithm sums the cost to reach that marking with a heuristic lower bound of the cost to the final marking, prioritizing promising paths until the optimal alignment is found. Despite $A^*$'s efficiency, finding an optimal alignment for long traces is computationally demanding. \modelspace aims to reduce the computational burden while maintaining high accuracy.

\section{Alignment of Long Event Sequences}\label{sec:method}

The proposed solution bounds the computational effort by partitioning a trace into {\em subtraces}, each aligned with respect to the model. For the decomposition, we define the notions of a {\em subtrace model} and {\em partial optimal alignment}.

\begin{definition}[Subtrace Model] \label{def:subtrace_model}
Let $A \subseteq \mathcal{A}$ be a set of activities over the universe of all possible activities $\mathcal{A}$, and $\sigma \in A^*$ be a sequence from the set of all possible activity sequences. Given a trace model $TN =(P,T,F,\lambda,m_i,m_f)$, a subtrace model $STN =(P_s,T_s,F_s,\lambda_s,m_{is},m_{fs})$ is a system net for a subsequence $\sigma_s = \sigma(j:k)$ for some $0 \leq j < k \leq |\sigma|$, such that $P_s = \{p_j,...,p_k\}$ is the set of places, $T_s \subseteq \{t_{j+1},...,t_k\}$ is the set of transitions, $F_s = \{(p_i,t_{i+1}) | j \leq i < k\} \cup \{(t_i,p_i)| j+1 \leq i \leq k\}$ is the set of flows, and $m_{is} = [p_j]$ and $m_{fs}=[p_k]$, are the initial and final markings, respectively. It holds that $\lambda_s(t_i)=\sigma(i)$ for $j+1 \leq i \leq k$, where $\lambda_s$ is a labeling function that assigns an activity label to each transition in the subtrace, mirroring the labeling function $\lambda$ in the full trace model while restricting it to subsequence $\sigma_s$.
\end{definition}

Standard alignment-based conformance checking seeks an optimal alignment for the entire synchronous product, while \modelspace aims to find the best alignment for each subtrace \textit{in the context of the whole process}. This technique adapts to varying process states in subtraces, starting with an initial state merging the subtrace's beginning with an applicable state of the process model and aiming to minimize deviation costs. It ends with a final state combining the subtrace's end with any legitimate final state of the model. \modelspace streamlines conformance checking by finding partial optimal alignments for each subtrace sequentially. 

\begin{definition}[Partial Optimal Alignment] \label{def:partial_opt_alignment}
Let $A \subseteq \mathcal{A}$ be a set of activities and $\sigma_s \in A^*$ a subtrace within a complete trace $\sigma$, with $STN =(P_s,T_s,F_s,\lambda_s,m_{is},m_{fs})$ as its corresponding subtrace model. Let $SN$ be a process model, and $SP_s$ the synchronous product of $SN$ and $STN$, reflecting the interaction between the process and the subtrace models. Let $c : T \rightarrow \mathbb{R}^+ \cup \{0\}$ be a cost function. A partial optimal alignment $\gamma^{opt}_s \in L_{SP_s}$ is the lowest cost execution sequence, starting from an initial marking $m_{init}$, which is the union of the initial marking $m_{is}$ in the $STN$ and a specific marking $m'$ of the $SN$, to a set of final markings. Each final marking is the union of the final marking $m_{fs}$ from $STN$ and any valid marking $m''$ within $SN$.
For all $\gamma_s \in L_{SP_s}$, it holds that $c(\gamma_s) \geq c(\gamma^{opt}_s)$, where $c(\gamma_s) = \sum_{1\leq i \leq  |\gamma_s|} \, c(\gamma_s(i))$. 
\end{definition}

Next, we introduce the overall approach of \modelspace for conformance checking over long traces, followed by illustrating the cost calculation and the sliding window mechanism. The pseudocode is given in Algorithm~\ref{alg:WindowBasedConformanceChecking}.

\subsection{Overall Approach}\label{sec:approach}

A long trace is sliced into subtraces (see Definition \ref{def:subtrace_model}) and the algorithm advances by iteratively sliding a window from one subtrace to the next. The effect of window length on performance and accuracy is analyzed in subsequent sections.

Decomposing the conformance checking problem into smaller subproblems is just one facet of our approach. To fully capitalize on this division and achieve an effective balance between computational efficiency and alignment accuracy, we couple the sliding window mechanism with an iterative alignment process that identifies situations in which an alignment that is optimal for the local window may lead to high future costs. In such cases, we select a different alignment. This strategy addresses the challenges of aligning long, complex traces while ensuring that the solution remains flexible and adaptable to the specific needs of the user.

\subsection{Local and Global Information}\label{sec:heuristic}

To improve the alignment process, we use both local and global information. By analyzing the process model, \modelspace identifies reachable and mandatory transitions from each marking \cite{casas2024reach}. This understanding helps establish a lower bound on the additional cost beyond the local window's conformance cost. 

For example, we can calculate (during a preprocessing step or on-the-fly) the number of activities remaining in the trace that cannot be executed from each state in the model. These unreachable activities translate into nonsynchronous moves, which contribute to a marginal future cost. This cost can be weighted based on the frequency of unreachable transitions within the remaining portion of the trace. Conversely, we can identify mandatory model transitions that must be executed to reach the final marking. Such mandatory transitions with labels that do not appear in the remaining trace are bound to generate nonsynchronous moves since there are no corresponding transitions with similar labels in the trace.
By considering these costs, \modelspace improves search accuracy and efficiency by discarding less promising alignments.

Consider a simple example of the process model shown in Figure \ref{fig:processmodelrunningexample} and the trace model in Figure \ref{fig:tracemodelrunningexample}. Assume that the process model is in marking $[p_{3}]$ (a single token in place $p_{3}$), and the marking within the trace model is $[p'_{5}]$. Thus, the marking corresponding to their synchronous product is $[p_{3}, p'_{5}]$.
One option is to perform a synchronous move \((E,E)\) with zero cost, leading to the marking \([p_4, p'_6]\). In this case, the algorithm would return a lower bound of 3 for the marginal cost because from \(p_4\), three transitions \((C, C, E)\) must be executed as nonsynchronous moves, and no mandatory transitions remain in the model.

A second, preferred, option is to execute the model transition \(\tau\), resulting in the marking \([p_2, p'_5]\) with $\epsilon$ cost and a lower bound marginal cost of 0. This is because the remaining transitions \((E, C, C, E)\) are reachable, and both mandatory transitions in the model \((C, E)\) are present in the trace. %As will be shown later, the second option is preferable.

\subsection{Iterative Alignment}\label{sec:sliding}

Algorithm~\ref{alg:WindowBasedConformanceChecking} begins by decomposing a trace into smaller subtraces (\hyperref[algline:Split]{line~1}). A detailed discussion of the hyperparameters that influence the trade-off between conformance checking accuracy and computational efficiency is presented in Section~\ref{sec:empirical_evaluation}.

The algorithm begins by initializing two lists: one to track visited markings and another to store computed alignments (\hyperref[algline:InitMarkings]{lines~2--3}). It then preprocesses the reachable and mandatory transitions for each model marking (\hyperref[algline:precalc]{line~4}). Each subtrace is processed sequentially by constructing a synchronous product with the process model to determine the alignments (\hyperref[algline:sync_product]{lines~5--6}).

For each subtrace except the last (see \hyperref[algline:extendedalign_init]{lines~\ref*{algline:extendedalign_init}--\ref*{algline:update_currentmarkings}}), the algorithm iteratively explores the reachability graph ($RG$) of the subtrace and process model to identify suitable alignments, considering final markings in the subtrace and any state in the process model as endpoints. Specifically, \hyperref[algline:heuristic_cost]{line~\ref*{algline:heuristic_cost}} extends $\alpha$ with the alignment from $RG_i$ that minimizes the total cost (the sum of the alignment cost and the cost of unreachable transitions remaining in the trace), leading to a unique final marking. After each alignment, it captures the final state and sets it as the starting point for the next subtrace, ensuring continuous conformance checking. The alignments are aggregated into a list.

For the last subtrace (see \hyperref[algline:finalalign_init]{lines~7--11}), the algorithm creates a \textit{FinalAlignments} list for complete alignments. It extends each alignment in the \textit{TopNAlignments} list from its most recently recorded state. These extended alignments are sorted by total cost, retaining the top $N_c$ lowest-cost alignments and their markings (typically, we select $N_c=2$ or $3$). Finally, the algorithm selects the alignment with the lowest total cost from the \textit{FinalAlignments} list as the optimal alignment, concluding its execution.

\subsection{Illustrative Example of \modelspace Execution} %Algorithm~\ref{alg:WindowBasedConformanceChecking} Execution}
\label{sec:illustrativeexample}

We illustrate the algorithm using the process and trace in Figures~\ref{fig:processmodelrunningexample} and \ref{fig:tracemodelrunningexample}, respectively. The model comprises six transitions, including a silent ($\tau$) one. For simplicity, we exclude the $\epsilon$ cost associated with silent transitions from our calculations. The trace $\langle ABDCCECCE \rangle$, depicted in the figure as a trace model, includes a sequence of 9 transitions. We set the hyperparameters to 2 alignments per subtrace and a window length of 3, allowing us to demonstrate at least two subtraces plus a final one.

The trace is split into three subtraces: $\langle ABD \rangle$, $\langle CCE \rangle$, and $\langle CCE \rangle$. Arrays are prepared to cache intermediate alignment candidates, and reachable transitions from each model marking are computed. For example, the reachable transitions from $[p_{0}]$ are ${A, B, C, D, E}$. The algorithm computes the two best alignments for the first subtrace $\langle ABD \rangle$, starting from $[p_{0}, p'_0]$ and ending at any model marking, that is, $[p_{2}, p'_3]$.

\begin{algorithm}
  \caption{\model: Window-Based Conformance Checking}
  \label{alg:WindowBasedConformanceChecking}
  \begin{algorithmic}[1]
    \Input{Process model $SN$, Trace $T$, number of candidate alignments per subtrace $N_c$, window length $L$.}
    \Output{Alignment between $T$ and $SN$.}
    
    \State Split trace $T$ into subtraces $t_1, t_2, \ldots$, each of length $L$. \label{algline:Split}
    \State Initialize $CurrentMarkings$ with $SN$'s initial marking. \label{algline:InitMarkings}
    \State Initialize $TopNAlignments$ to hold $N_c$ empty alignments. \label{algline:InitAlignments}
    \State Pre-calculate reachable and mandatory transitions for each marking in the process model. \label{algline:precalc}
    
    \For{$i = 1, 2, \dots, \left\lceil \frac{|T|}{L} \right\rceil$}
      \State Construct synchronous product $SP_i$ from $t_i$ and $SN$. \label{algline:sync_product}
      \If{$i = \left\lceil \frac{|T|}{L} \right\rceil$}
        \State Initialize $FinalAlignments$ as an empty list. \label{algline:finalalign_init}
        \ForAll{alignment $\alpha$ in $TopNAlignments$}
          \State Extend $\alpha$ to include alignment from $RG_{\left\lceil \frac{|T|}{L} \right\rceil}$ 
          \Statex \hspace{1.35cm} starting from $\alpha$'s final marking.
          \State Add this extended alignment to $FinalAlignments$.
        \EndFor
        \State \textbf{break}
      \Else
        \State Initialize $ExtendedAlignments$ as an empty list. \label{algline:extendedalign_init} % <-- New label at start of block
        \State Initialize $UniqueFinalMarkings$ as an empty set.
        \ForAll{alignment $\alpha$ in $TopNAlignments$}
          \For{$j = 1$ to $N_c$}
            \State Extend $\alpha$ with the alignment computed for
            \Statex \hspace{1.85cm} subtrace $i$ and $\alpha$'s previous final marking. 
            \label{algline:heuristic_cost}
            \State Add extension to $ExtendedAlignments$.
            \State Add final marking to $UniqueFinalMarkings$.
          \EndFor
        \EndFor
        \State Sort $ExtendedAlignments$ by cost and keep top $N_c$.
        \State Update $TopNAlignments$ with top $N_c$ alignments.
        \State Update $CurrentMarkings$ with final markings. \label{algline:update_currentmarkings} % <-- New label at end of block
      \EndIf
    \EndFor
    
    \State \Return{Lowest cost alignment from $FinalAlignments$.}
  \end{algorithmic}
\end{algorithm}

Alignments in Figures~\ref{subfig:table1} and~\ref{subfig:table2} correspond to the markings $[p_{2}, p'_3]$ and $[p_{0}, p'_3]$, each incurring a cost of 1 due to a nonsynchronous move. The remaining transitions labeled $C$ and $E$ can be reached from both $[p_2]$ and $[p_0]$. However, the mandatory transitions $(A, B)$ from $[p_0]$ do not appear within the leftover portion of the trace, resulting in a lower bound marginal cost of 0 and 2, respectively. Both alignments and their final markings are recorded for further processing.

The algorithm then aligns the second subtrace $\langle CCE \rangle$, starting from the final markings of the first subtrace: $[p_{2}, p'_3]$ and $[p_{0}, p'_3]$. This results in two unique optimal alignments from each starting point, as shown in Figures~\ref{subfig:table3} and~\ref{subfig:table5} (from $[p_{2}, p'_3]$) and Figures~\ref{subfig:table4} and~\ref{subfig:table6} (from $[p_{0},p'_3]$). 

The best two alignments from $[p_{2}, p'_3]$ incur a cost of 1 each, which leads to a total cost of 2. Although there is a perfect local alignment with a cost of 0 (ending at $[p_{4}, p'_6]$), it is disqualified as the lower bound marginal cost is 3 due to unreachable transitions in the final subtrace from $[p_4]$. Alignments from $[p_{0}, p'_3]$ accumulate a total cost of 4. The algorithm retains the two lowest-cost alignments with final markings of $[p_{2}, p'_6]$ and $[p_{3}, p'_6]$, both originating from $[p_{2}, p'_3]$. None of these alignments include activity $E$ as a synchronous move due to the marginal cost, as firing activity $E$ in the process model would require additional nonsynchronous moves. This increases the marginal costs.

\begin{figure}[ht!]
    \centering
    % Table 1
    \begin{subfigure}{\columnwidth}
        \begin{adjustbox}{valign=t}
        \begin{minipage}{\columnwidth}
            \centering
            \begin{tabular}{c|c|c|c}
                log & $A$ & $B$ & $D$  \\
                \hline
                model & $A$ & $B$ & $\gg$ 
            \end{tabular}\\[5pt]
            Marking: $[p_{2}, p'_3]$, Accumulated cost: 1
        \end{minipage}
        \end{adjustbox}
        \caption{Alignment for first subtrace ending at $[p_{2}, p'_3]$.}
        \label{subfig:table1}
    \end{subfigure}

    \vspace{1pt} % Adjust vertical space to separate the groupings visually

    % Table 2
    \begin{subfigure}{\columnwidth}
        \begin{adjustbox}{valign=t}
        \begin{minipage}{\columnwidth}
            \centering
            \begin{tabular}{c|c|c|c|c}
                log & $A$ & $B$ & $\gg$ & $D$  \\
                \hline
                model & $A$ & $B$ & $C$ & $D$ 
            \end{tabular}\\[5pt]
            Marking: $[p_{0}, p'_3]$, Accumulated cost: 1
        \end{minipage}
        \end{adjustbox}
        \caption{Alternative alignment for first subtrace ending at $[p_{0}, p'_3]$.}
        \label{subfig:table2}
    \end{subfigure}

    \vspace{1pt} % Adjust vertical space between the groupings

    % Table 3
    \begin{subfigure}{\columnwidth}
        \begin{adjustbox}{valign=t}
        \begin{minipage}{\columnwidth}
            \centering
            \begin{tabular}{c|c|c|c|c|c}
                log & $C$ & $\gg$ & $C$ & $\gg$ & $E$  \\
                \hline
                model & $C$ & $\tau$ & $C$ & $\tau$ & $\gg$ 
            \end{tabular}\\[5pt]
            Marking: $[p_{2}, p'_6]$, Accumulated cost: 2
        \end{minipage}
        \end{adjustbox}
        \caption{Alignment from $[p_{2}, p'_3]$ to $[p_{2}, p'_6]$.}
        \label{subfig:table3}
    \end{subfigure}

    \vspace{1pt} % Adjust vertical space between the groupings

    % Table 4
    \begin{subfigure}{\columnwidth}
        \begin{adjustbox}{valign=t}
        \begin{minipage}{\columnwidth}
            \centering
            \begin{tabular}{c|c|c|c|c|c|c}
                log & $\gg$ & $\gg$ & $C$ & $\gg$ & $C$ & $E$  \\
                \hline
                model & $A$ & $B$ & $C$ & $\tau$ & $C$ & $\gg$ 
            \end{tabular}\\[5pt]
            Marking: $[p_{2},p'_6]$, Accumulated cost: 4
        \end{minipage}
        \end{adjustbox}
        \caption{Suboptimal alignment from $[p_{0}, p'_3]$ to $[p_{2}, p'_6]$.}
        \label{subfig:table4}
    \end{subfigure}

    \vspace{1pt} % Adjust vertical space between the groupings

    % Table 5
    \begin{subfigure}{\columnwidth}
        \begin{adjustbox}{valign=t}
        \begin{minipage}{\columnwidth}
            \centering
            \begin{tabular}{c|c|c|c|c}
                log & $C$ & $\gg$ & $C$ & $E$  \\
                \hline
                model & $C$ & $\tau$ & $C$ & $\gg$ 
            \end{tabular}\\[5pt]
            Marking: $[p_{3},p'_6]$, Accumulated cost: 2
        \end{minipage}
        \end{adjustbox}
        \caption{Alignment from $[p_{2}, p'_3]$ to $[p_{3}, p'_6]$.}
        \label{subfig:table5}
    \end{subfigure}

    \vspace{1pt} % Adjust vertical space between the groupings

    % Table 6
    \begin{subfigure}{\columnwidth}
        \begin{adjustbox}{valign=t}
        \begin{minipage}{\columnwidth}
            \centering
            \begin{tabular}{c|c|c|c|c|c|c}
                log & $\gg$ & $\gg$ & $C$ & $\gg$ & $C$ & $E$  \\
                \hline
                model & $A$ & $B$ & $C$ & $\tau$ & $C$ & $\gg$ 
            \end{tabular}\\[5pt]
            Marking: $[p_{3},p'_6]$, Accumulated cost: 4
        \end{minipage}
        \end{adjustbox}
        \caption{Suboptimal alignment from $[p_{0}, p'_3]$ to $[p_{3}, p'_6]$.}
        \label{subfig:table6}
    \end{subfigure}

    \vspace{1pt} % Adjust vertical space between the groupings

    % Table 7
    \begin{subfigure}{\columnwidth}
        \begin{adjustbox}{valign=t}
        \begin{minipage}{\columnwidth}
            \centering
            \begin{tabular}{c|c|c|c|c|c}
                log & $\gg$ & $C$ & $\gg$ &$C$ & $E$ \\
                \hline
                model &$\tau$ & $C$ & $\tau$ & $C$ & $E$
            \end{tabular}\\[5pt]
            Marking: $[p_{4},p'_9]$, Accumulated cost: 2
        \end{minipage}
        \end{adjustbox}
        \caption{Alignment from $[p_{3}, p'_6]$ to $[p_{4}, p'_9]$.}
        \label{subfig:table7}
    \end{subfigure}

    \vspace{1pt} % Adjust vertical space between the groupings

    % Table 8
    \begin{subfigure}{\columnwidth}
        \begin{adjustbox}{valign=t}
        \begin{minipage}{\columnwidth}
            \centering
            \footnotesize
            \begin{tabular}{c|c|c|c|c}
                log & $C$ & $\gg$ & $C$ & $E$ \\
                \hline
                model & $C$ & $\tau$ & $C$ & $E$
            \end{tabular}\\%[5pt]
            Marking: $[p_{4},p'_9]$, Accumulated cost: 2
        \end{minipage}            
        \end{adjustbox}
        \caption{Optimal alignment from $[p_{2}, p'_6]$ to $[p_{4}, p'_9]$.}
        \label{subfig:table8}
    \end{subfigure}

%\Description{Markings and sub-alignment costs for subtraces.}
\caption{Markings and alignment costs for subtraces. (a) Alignment for first subtrace ending at $[p_{2}, p'_3]$. (b) Alternative alignment for first subtrace ending at $[p_{0}, p'_3]$. (c) Alignment from $[p_{2}, p'_3]$ to $[p_{2}, p'_6]$. (d) Suboptimal alignment from $[p_{0}, p'_3]$ to $[p_{2}, p'_6]$. (e) Alignment from $[p_{2}, p'_3]$ to $[p_{3}, p'_6]$. (f) Suboptimal alignment from $[p_{0}, p'_3]$ to $[p_{3}, p'_6]$. (g) Alignment from $[p_{3}, p'_6]$ to $[p_{4}, p'_9]$. (h) Optimal alignment from $[p_{2}, p'_6]$ to $[p_{4}, p'_9]$.} 
\label{fig:subalignments}
\end{figure}

The algorithm then aligns the last subtrace $\langle CCE \rangle$, starting from markings $[p_{2}, p'_6]$ and $[p_{3}, p'_6]$. The search focuses on a single optimal alignment for each starting marking, reflecting the final integrated marking of both the trace and model. Both alignments accumulate a total cost of 2, as shown in Figures~\ref{subfig:table7} and~\ref{subfig:table8}. 

Since both alignments have the same cost, one is chosen arbitrarily (in this case, the alignment from $[p_{2}, p'_6]$). The final output is the combined sequence of alignments, leading to the optimal alignment with a conformance cost of 2.

\section{Complexity Analysis}\label{sec:complexity}

In this section, we analyze the computational complexity of \model. The key idea is to partition a trace of length \(N\) into subtraces (or windows) of fixed length \(L\), thereby confining the exponential search cost to a small window rather than to the entire trace.

\subsection{Per-Window Alignment Complexity}

Let the full trace be of length \(N\). We partition it into
\[
W = \lceil N / L \rceil
\]
subtraces, where each subtrace (except possibly the last) contains \(L\) events.
For each subtrace, we compute an alignment over the synchronous product of the subtrace and the process model. In alignment-based conformance checking, the worst-case search cost grows exponentially with the search depth. 

In our approach, consider an alignment for an \(L\)-length subtrace. If every event in the subtrace incurred a nonsynchronous move, the maximum possible cost would be at most \(L\). To ensure that the alignment remains competitive (i.e., does not exceed the cost of a fully non-matching subtrace), the longest possible alignment may consist of \(L\) nonsynchronous model moves followed by \(L\) synchronous moves, leading to a total alignment length of \(2L\). 

Additionally, a limited number of silent moves (since they cost $\epsilon$ and each marking is expanded at most once) may be included to account for internal flow logic within the process model. As a result, we bound the effective search depth per subtrace as:
\[
d \approx 2L + S,
\]
where \(S\) is a constant determined by the structure of the process model, representing the maximal number of extra silent moves.

Let \(b\) denote the worst-case branching factor. In practice, \modelspace reduces the number of explored paths, so we denote the effective branching factor by \(b_{\text{eff}}\) (with \(b_{\text{eff}} \leq b\)). Moreover, we maintain a small fixed number (see, $N_c$ in Algorithm~\ref{alg:WindowBasedConformanceChecking}) of candidate alignments per subtrace. Hence, the time complexity to align one subtrace is
\[
O\!\Bigl( N_c \cdot b_{\text{eff}}^{\,d} \Bigr)
\quad \text{or} \quad
O\!\Bigl( N_c \cdot b_{\text{eff}}^{\,2L + S} \Bigr).
\]

\subsection{Overall Time Complexity}

We process \(W = \lceil N / L \rceil\) subtraces sequentially. As a result, the overall time complexity is the sum of the costs for all windows, plus an additional term to account for processing a short ``tail" alignment at the end. This tail refers to the leftover transitions within the process model needed to reach its final marking. Let \(L_m\) represent the constant number of additional moves required to handle this tail.
 The overall worst-case time complexity is then
\[
O\!\Bigl( \lceil N / L \rceil \cdot N_c \cdot b_{\text{eff}}^{\,2L + S} + b^{L_m} \Bigr).
\]
In this expression:
\begin{description}[leftmargin=*]
    \item[Exponential term:] The term \(b_{\text{eff}}^{\,2L + S}\) depends on the window size \(L\) (and a constant \(S\)) rather than the to full trace length \(N\). Since \(L \ll N\) in typical applications, the exponential increase is effectively bounded.
    \item[Candidate alignments:] The constant \(N_c\) is small, so its impact on the overall complexity is minimal.
    \item[Tail alignment:] The term \(b^{L_m}\) (with a constant $L_m\ll N$) accounts for the tail alignment and does not affect scalability with respect to \(N\).
\end{description}

\subsection{Space Complexity}

The space complexity per window depends on the number of nodes stored during the search. Since candidate alignments are evaluated sequentially, and only the resulting alignment (along with its cost) is retained—rather than the entire search tree or all candidate nodes—the space complexity for a single window is bounded by:

\[
O\Bigl( b_{\text{eff}}^{\,2L + S} \Bigr).
\]

In the final window, any remaining transitions in the process model must be executed to reach its final marking. The alignment of these transitions requires space proportional to:

\[
O\Bigl( b^{L_m} \Bigr),
\]

where \(L_m\) is the previously defined number of moves required to complete the alignment.

Since the algorithm processes windows sequentially, at any given time, it only needs to store either the candidate alignment for the current window or, in the case of the final window, the alignment incorporating the remaining transitions. As a result, the overall space complexity is determined by the larger of these two terms:

\[
O\Bigl(\max\bigl(b_{\text{eff}}^{\,2L+S},\, b^{L_m}\bigr)\Bigr).
\]

\subsection{Summary and Practical Implications}

By partitioning a trace of length \(N\) into \(W = \lceil N / L \rceil\) windows, our approach achieves an overall worst-case time complexity of
\[
O\!\Bigl( \lceil N / L \rceil \cdot N_c \cdot b_{\text{eff}}^{\,2L + S} + b^{L_m} \Bigr).
\]
Because \(N_c\), \(S\), and \(L_m\) are constants and \(L\) is chosen such that \(L \ll N\), the exponential component is limited to a manageable window size. In practice, \modelspace significantly reduces the effective branching factor \( b_{\text{eff}} \), leading to substantially better performance than the worst-case bound would suggest. This reduction transforms the problem from an intractable \( O(b^N) \) complexity (for full-trace alignment) to an exponential dependence on \( L \), making alignment-based conformance checking feasible even for traces containing thousands of events. In Section~\ref{sec:empirical_evaluation}, we provide empirical evidence supporting this theoretical analysis.

\section{Empirical Evaluation}\label{sec:empirical_evaluation}
We conducted an extensive experimental study to assess the performance and scalability of \modelspace compared to state-of-the-art optimal and heuristic conformance checking approaches. Our evaluation was structured around three distinct settings:

\textbf{Classic process mining datasets:} These datasets contain short to moderate-length traces, with a reference process model manually discovered from the dataset. They serve to investigate the accuracy of our approach. Regarding speed, our approach is designed for longer traces. In the case of short to moderate traces, the overhead of decomposing a trace into smaller subtraces--each explored individually--is expected to be less time-efficient compared to performing conformance checking over the complete trace.

\textbf{Challenging process mining datasets:} These datasets provide, for each log, a predefined process model known to poorly fit the log. They serve to investigate the accuracy of our approach in challenging settings.
    
\textbf{Procedural computer vision datasets of food preparation:} These datasets feature extremely long traces, highlighting the need for a scalable and accurate conformance checking approach. They serve to explore both the scalability and accuracy of \modelspace compared to leading alternatives.

Optimal algorithms and \modelspace were implemented in Python, while heuristic approaches ran in Java using their original code. Although Java is generally faster, placing our algorithm at a disadvantage in CPU performance, \modelspace still outperforms for long event sequences, demonstrating its efficiency. The experiments were run in Docker containers on an Intel Xeon E5-2650 @2.20 GHz machine (24 cores, 2 threads per core), with each container allocated 10 CPUs and 30 GB of memory.

\modelspace was compared with:

\textbf{Optimal methods:} Standard $A^*$ search \cite{adriansyah2014aligning}, REACH \cite{casas2024reach} and $A^*$ Incremental \cite{van2018efficiently} (the latter consistently timed out and was therefore excluded from the results.)

\textbf{Heuristic approaches:} The Trie-based method \cite{awad2021efficient} and the Tandem Repeats (TR) approach \cite{reissner2020efficient} were selected since both have been shown to outperform other heuristic methods.

The evaluation is conducted through three experiments, with the results summarized in Tables~\ref{tab:results-part1}, \ref{tab:results-part2-delta}, and \ref{tab:food-extended}. In all experiments, the reported time, cost and percentages represent the average across all traces in the log.

\subsection{Classic Datasets}\label{sec:classic}

Table~\ref{tab:results-part1} presents the performance evaluation on the Sepsis and BPIC datasets. We filtered out traces with fewer than 80 events to ensure that only short to moderate traces remain, as shorter traces can be efficiently processed by existing approaches.

For each dataset, we derived a process model using the Inductive Miner~\cite{leemans2014discovering}, trained on 10 randomly selected traces to balance fitness and precision~\cite{bogdanov2023sktr}. We applied \modelspace with small window sizes ranging from 5 to 50 events. These small windows that increased the computational burden by increasing the number of synchronous product constructions were selected to model traces that significantly exceed the window size (otherwise, \modelspace could be much faster).
For a fair comparison with the Trie-based approach~\cite{awad2021efficient}, we followed its methodology, using the same percentage of traces from the filtered event log to construct the Trie data structure.

As shown in Table~\ref{tab:results-part1}, \modelspace achieves optimal solutions while reducing execution times. Trie-based approach achieves faster execution on BPIC 2017 but fails to produce optimal alignments. For these short to moderate trace lengths, TR maintains optimality and delivers fast computation times. As we show later, TR’s performance deteriorates significantly for long traces.

\subsection{Evaluation on Poorly Fitting Models}\label{sec:noise}

To assess robustness, we evaluated \modelspace on the pr. datasets~\cite{van2018efficiently}, which contain reference models deliberately designed to poorly capture the log behavior. This setting increases conformance checking complexity due to high misalignment between the log and model.

While \modelspace is well suited for these datasets, the Trie-based method relies on a log-derived data structure, making it unsuitable for conformance checking when using predefined process models. Similarly, the TR approach is limited to state machine workflow Petri nets, where each transition has a single input and output place. As a result, it cannot handle reference models without significant modifications. Since \modelspace is the only heuristic capable of handling these datasets, we compared it to the optimal approaches.

Table~\ref{tab:results-part2-delta} presents the results for these datasets. \modelspace consistently outperforms competing methods in execution speed, achieving up to a 10× speedup while computing optimal alignments in most cases. When not fully optimal, the deviation (i.e., $\Delta$ cost) remains minimal, averaging just 0.6\% from the optimal cost. These results demonstrate that even in highly misaligned scenarios, \modelspace effectively balances efficiency and accuracy, delivering substantial speed improvements while maintaining near-optimal alignment quality.

\subsection{Scalability on Very Long Traces}\label{sec:food_datasets}

As noted, \modelspace was designed to efficiently handle very long traces which the focus of the next experiment. For evaluating its performance, we use three food preparation datasets--GTEA, Breakfast, and 50 Salads~\cite{DatasetLink}--which contain traces with thousands of events, presenting challenges due to their length and variability.

For model discovery, we sampled 10 traces per dataset, except for Breakfast, where 15 traces were used due to its greater diversity and higher number of unique trace variants. A 120-second timeout per trace was imposed to ensure a fair evaluation.

While we tested all approaches, from the optimal ones only REACH could compute alignments for some traces within the time limit, thus they are excluded. Table~\ref{tab:food-extended} reports the percentage of traces successfully processed by REACH and the Trie-based approach. Since neither method completed alignments for all traces in any dataset, their average alignment costs and computation times were omitted (both were far worse compared to \modelspace). The TR approach aligned all traces in GTEA and Breakfast but could not handle the 50 Salads dataset within the time limit. Moreover, its conformance costs exceeded those of \modelspace as can be seen in the table.

Among all methods, \modelspace was the only one to process all traces within the allocated time while also achieving lower conformance costs when comparisons were possible. These results underscore its robustness and scalability, as it efficiently handles long event sequences while maintaining competitive alignment costs, even in highly challenging real-world scenarios.

%In contrast, REACH and the Trie-based approach completed only a subset of traces. While TR executed successfully on GTEA and Breakfast, it incurred significantly higher alignment costs than \modelspace. The process model for Breakfast exhibited low fitness due to the high trace variability, leading to costly alignments as many trace variants were not well represented.

%These results highlight the robustness and scalability of \model: it efficiently handles long event sequences while maintaining competitive alignment costs, even in highly challenging real-world scenarios.

\subsection{Window Length vs. Performance}\label{sec:window_analysis}

We conducted a sensitivity analysis to examine the trade-off between computational efficiency and alignment quality concerning the sliding window length $L$. This trade-off arises from \model's use of a sliding window to decompose traces into subtraces. As the window length increases, fewer subtraces are needed, potentially improving conformance accuracy; however, computation time per subtrace also increases.

\begin{figure}[h!]
    \centering
    \includegraphics[width=1.0\columnwidth]{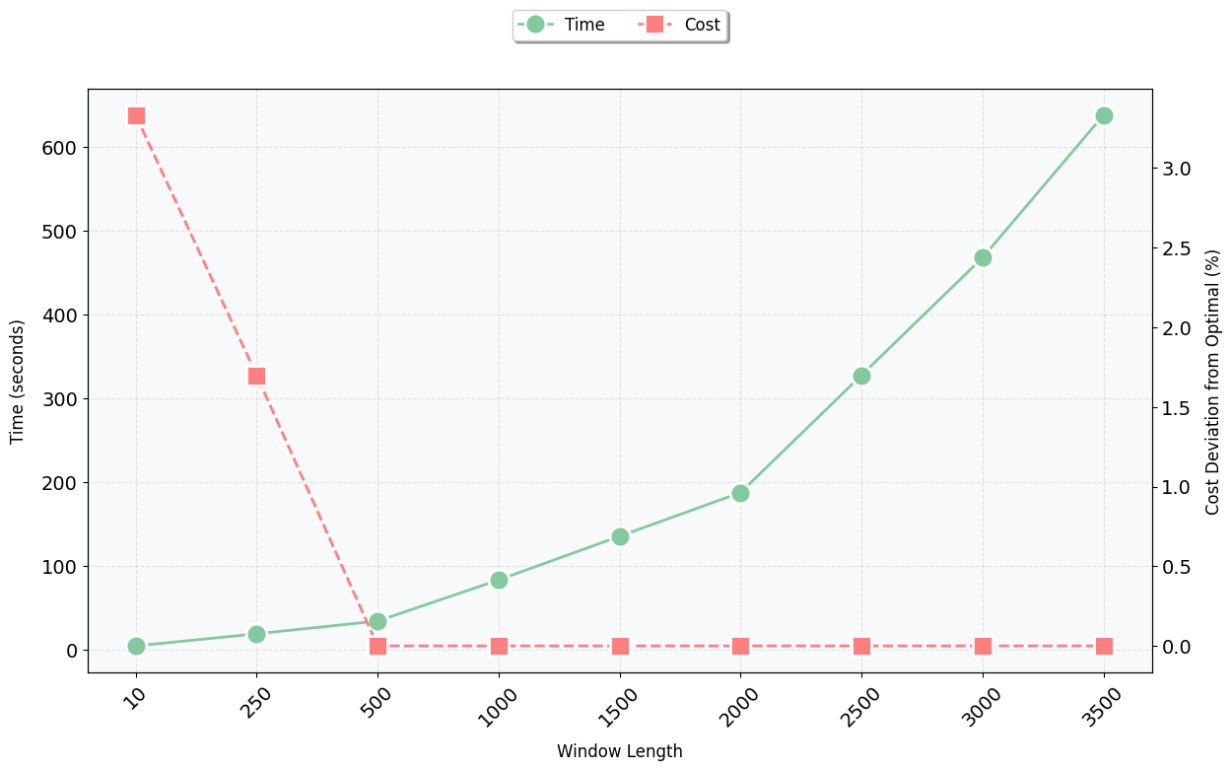}
    %\Description{A graph comparing alignment time and cost deviation versus window length $L$ for \model. As the window length $L$ increases, the number of subtraces decreases, but the computation time per subtrace increases. Larger window lengths ensure that the cost converges to the optimal value, with $L=3500$ representing full trace processing.}
    \caption{\model's~ alignment time and cost deviation versus window length $L$. As $L$ increases, fewer subtraces are required, but computation time per subtrace grows. Larger $L$ ensures cost convergence to the optimal, with $L=3500$ representing full trace processing.}
    \label{fig:complexityVsWindow}
\end{figure}

Figure~\ref{fig:complexityVsWindow} illustrates that increasing the window length \(L\) results in longer overall computation times while simultaneously reducing cost deviation, thereby enhancing conformance accuracy. Notably, once \(L\) reaches 500, the computation time begins to grow exponentially; every additional 500 units in length increases the computation time by roughly 1.5 times. (For full trace processing, a window length of \(L = 3500\) is used.)

\subsection{Summary}\label{sec:empirical_summary}
In summary, \modelspace outperforms other leading methods when considering both time efficiency and accuracy. For standard datasets, it computes optimal alignments while achieving up to 10× speedups over REACH. On low-fitting models, such as pr-\textit{noise} datasets, it maintains near-optimal alignments with an average deviation of just 0.6\%. The other heuristic methods struggle with such datasets, where conformance is performed against a predefined process model--a common real-world requirement.

\modelspace is also the only method capable of processing all traces in datasets with extremely long sequences (e.g., food preparation datasets) within the given time constraints. Competing methods either failed entirely or incurred significantly higher alignment costs.

These results establish \modelspace as a highly efficient, robust, and scalable conformance checking solution for real-world scenarios involving long traces.

 %Our analysis of sliding window length confirms that increasing the window size improves accuracy, with a length of 500 ensuring optimal conformance while maintaining practical feasibility through heuristic efficiency.

% ---------------------- TABLES ----------------------

\begin{table*}[t]
    \centering
    \caption{Sepsis and BPIC Datasets: Comparison of Algorithm Performance.}
    \label{tab:results-part1}
    \resizebox{\textwidth}{!}{%
        \begin{tabular}{|c|c|c|c|c|c|c|c|c|c|c|}
            \hline
            \textbf{Dataset} & \textbf{Traces (\#)} & \textbf{Avg. Len} & \textbf{$A^*$ Time (s)} & \textbf{REACH (s)} & \textbf{Trie (s)} & \textbf{Opt. Trie (\%)} & \textbf{TR (s)} & \textbf{Opt. TR (\%)} & \textbf{\modelspace (s)} & \textbf{Opt. \modelspace (\%)} \\
            \hline
            Sepsis            & 5    & 129 & 23.3   & 0.85  & \textbf{0.01} & \textbf{100}   & \textbf{0.01} & \textbf{100}   & \textbf{0.01} & \textbf{100}  \\
            \hline
            BPIC 2012         & 181  & 98  & 143.9  & 4.77  & 1.43        & 35.6  & \textbf{0.27} & \textbf{100}   & 1.24 & \textbf{100} \\
            \hline       
            BPIC 2017         & 799  & 94  & 41.85  & 3.55  & 0.7& 69.3  & \textbf{0.08} & \textbf{100}   & 1.38        & \textbf{100} \\
            \hline
            BPIC 2019         & 896  & 195 & 144.5  & 4.76  & 2.67        & 10.9  & \textbf{0.02} & \textbf{100}   & 1.34 & \textbf{100} \\
            \hline
        \end{tabular}%
    }
\end{table*}

\begin{table*}[t]
    \centering
    \caption{pr-Noise Datasets: Performance Evaluation Using A$^*$, REACH, and \modelspace (Optimal and $\Delta$ Cost in \%).}
    \label{tab:results-part2-delta}
     \begin{adjustbox}{width=\textwidth}
    \begin{tabular}{|l|c|c|c|c|c|c|c|}
        \hline
        \textbf{Dataset} & \textbf{Traces (\#)} & \textbf{Avg. Len} & \textbf{$A^*$ Time (s)} & \textbf{REACH (s)} & \textbf{\modelspace (s)} & \textbf{Opt. \modelspace (\%)} & \textbf{$\Delta$ Cost (\%)} \\
        \hline
        pr-A48-m37-noise  & 92   & 86  & 8.15  & 0.59  & \textbf{0.05} & 100   & 0 \\
        \hline
        pr-A48-m50-noise  & 422  & 99  & 14.58 & 1.08  & \textbf{0.25} & 90    & 1 \\
        \hline
        pr-A57-m39-noise  & 61   & 85  & 16.61 & 1.01  & \textbf{0.5}  & 96.2  & 0.4 \\
        \hline
        pr-A57-m52-noise  & 421  & 92  & 26.72 & 1.60  & \textbf{0.46} & 95.8  & 0.6 \\
        \hline
        pr-A59-m41-noise  & 169  & 85  & 12.73 & 1.03  & \textbf{0.29} & 100   & 0 \\
        \hline
        pr-A59-m55-noise  & 493  & 99  & 23    & 2.02  & \textbf{0.46} & 88.7  & 1.8 \\
        \hline
    \end{tabular}
    \end{adjustbox}
\end{table*}

\begin{table*}[t]
    \centering
    \caption{Food Datasets: Metrics for REACH, Trie, TR \& \modelspace (N/E: Not executed -- could not handle long sequences).}
    \label{tab:food-extended}
    \resizebox{\textwidth}{!}{%
        \begin{tabular}{|l|c|c|c|c|c|c|c|c|}
            \hline
            \textbf{Dataset} & \textbf{Traces (\#)} & \textbf{Avg. Len} & \textbf{REACH Completion (\%)} & \textbf{Trie Completion (\%)} & \textbf{TR Time (s)} & \textbf{TR Cost} & \textbf{\modelspace Time (s)} & \textbf{\modelspace Cost} \\
            \hline
            GTEA        & 28   & 1301  & 71.4\%   & 85\%   & \textbf{0.12}  & 143    & 3.3   & \textbf{0.1}  \\
            \hline
            Breakfast   & 1008 & 2005  & 19.6\%   & 0\%    & \textbf{0.01}  & 672.6  & 8.2   & \textbf{371.4}  \\
            \hline
            50 Salads   & 40   & 5945  & 0\%     & 0\%    & N/E   & N/E    & \textbf{98}    & \textbf{1.7}  \\
            \hline
        \end{tabular}%
    }
\end{table*}

\section{Related Work}
\label{sec:literature_review}
Enhancing the performance of conformance checking is a research area that intersects with planning, declarative modeling, compliance checking, and query checking~\cite{de2017disruptive, chiariello2022asp, alman2023monitoring}.

Several approaches were suggested to handle the computational challenges of alignment-based conformance checking. A method leveraging an extended marking equation to accelerate the search for optimal alignments~\cite{van2018efficiently} encounters exponential complexity as trace lengths increase. Similarly, \cite{casas2024reach} developed the REACH algorithm, which mitigates computational overhead by preprocessing mandatory transitions but still faces challenges with very long traces.

Approximation-based conformance checking is another significant line of research.  A foundational approach using subset selection and edit distance techniques selects proxy behavior from the model, providing theoretical bounds on conformance values~\cite{fani2020conformance}. Building on this work, \cite{awad2021efficient} proposed using Trie data structures to encode and efficiently search the proxy behavior space, achieving logarithmic search space reduction. However, their approach can show significant deviation from optimal alignments, particularly for traces that poorly conform to the process model. Another contribution by \cite{reissner2020efficient} explored handling tandem repeats in traces, though it specifically targets scenarios with long loops rather than addressing the general case.

%Other notable methodologies include 
Declarative approaches~\cite{de2017disruptive, chiariello2022asp}, which show promise but face scalability limitations, and prefix alignment-based techniques~\cite{schuster2022conformance} have also been proposed in the literature. Approaches utilizing recomposition~\cite{lee2018recomposing, lee2017replay} and sampling strategies~\cite{bauer2022sampling} provide rapid approximations yet do not explicitly compute alignments, limiting their interpretability usefulness. %for detailed conformance analysis.

Efficiently processing very long traces remains a persistent challenge. While recent works have made significant progress through approximation techniques \cite{fani2020conformance, awad2021efficient}, a comprehensive solution balancing efficiency and accuracy for arbitrary long traces is still needed. This work fills this gap by introducing a sliding window approach designed to handle extremely long traces efficiently. It achieves near-optimal conformance results within realistic computational times while making no assumptions about the trace structure. \modelspace bridges the performance-accuracy trade-off and advances the state-of-the-art in conformance checking for long event sequences.

\section{Conclusions and Future Directions}\label{sec:conclusions_and_future_directions}

We introduced \model, a sliding window conformance checking method that efficiently aligns long event sequences by decomposing traces into manageable subtraces. By leveraging both local and global structural information from the process and trace models, \modelspace confines exponential complexity to a small, user-defined window rather than the full trace. Our theoretical analysis and empirical evaluation demonstrate that \modelspace consistently computes near-optimal alignments while significantly reducing computation time, even for poorly fitting models or extremely long traces.

Several extensions can further enhance the approach. Adaptive window sizing techniques could dynamically adjust the window length based on trace characteristics, optimizing efficiency and accuracy. Integrating domain-specific heuristics or machine learning may refine cost estimations and improve pruning. Extending the framework to multi-dimensional or heterogeneous event logs would broaden its applicability, while parallel or distributed architectures could enable real-time conformance checking on large-scale datasets.

In summary, \modelspace is a robust and scalable solution for alignment-based conformance checking, with these future directions expected to further enhance its performance and applicability in process mining.

\newpage
% Use named bibliography style
\bibliographystyle{ACM-Reference-Format}
\bibliography{kdd}

\end{document}